\documentclass[a4paper]{jpconf}  
\usepackage{graphicx}
\newcommand{\be}{\begin{equation}}
\newcommand{\ee}{\end{equation}}  
\newcommand{\bea}{\begin{eqnarray}}
\newcommand{\eea}{\end{eqnarray}}
\newcommand{\noi}{\noindent}

\def\gsim{
    \mathrel{\rlap{\raise 0.511ex
        \hbox{$>$}}{\lower 0.511ex \hbox{$\sim$}}}}

\def\lsim{
    \mathrel{\rlap{\raise 0.511ex
        \hbox{$<$}}{\lower 0.511ex \hbox{$\sim$}}}}

\begin{document}

\title{Intense Neutrino Beams and Leptonic CP Violation}

\author{William~J.~Marciano and Zohreh Parsa}
 
\address{Brookhaven National Laboratory\footnote[1]{This work is 
supported by the U.S. Department of Energy under contract number 
DE-AC02-98-CH-10886.}Upton, New York\ \ 11973}\ead{marciano@bnl.gov, 
parsa@bnl.gov}
      
\begin{abstract}
Effects of the Leptonic CP violating phase, $\delta$, on 3 
generation neutrino oscillation rates and asymmetries are 
discussed. A figure of merit argument is used to show that
our ability to measure the phase $\delta$ is rather insensitive 
to the  value of $\theta_{13}$ (for $\sin^22\theta_{13}\gsim0.01$) 
as well as the detector distance (for very long oscillation baselines). 
Using a study of   $\nu_\mu\to\nu_e$ oscillations for BNL-Homestake 
(2540 km) we show that a conventional horn focused wide band neutrino 
beam generated by an intense 1-2 MW proton source combined with a very large 
water Cherenkov detector (250-500 kton) should be able to 
determine  $\delta$ to about $\pm 15^\circ$ in $5 \times 10^7 sec.$ of
running.  In addition, such an effort would also measure the other 
oscillation parameters   ($\theta_{ij}$, $\Delta m^2_{ij}$) with 
high precision.  Similar findings apply to a Fermilab-Homestake (1280 km)
baseline.  We also briefly discuss features of Superbeams, Neutrino Factories
and Beta-Beams. 

\end{abstract}

\section{Status Of 3 Generation Lepton Mixing}

The known weak interaction states~ $|\nu_\ell>,~ \ell=e,\mu,~\tau$
produced in charged current interactions are related to the neutrino
mass eigenstates $|\nu_i>, i=1,2,3$ with masses $m_i$ by the $3\times3$
unitary matrix $U$.

{
\be
\pmatrix{|\nu_e> \cr |\nu_\mu> \cr |\nu_\tau>} = U \pmatrix{|\nu_1> \cr
|\nu_2> \cr |\nu_3>}
\ee

\[
U = \pmatrix{ c_{12}c_{13} & s_{12}c_{13} & s_{13}e^{-i\delta} \cr
-s_{12}c_{23}-c_{12}s_{23}s_{13}e^{i\delta} &
c_{12}c_{23}-s_{12}s_{23}s_{13}e^{i\delta} & s_{23}c_{13} \cr
s_{12}s_{23}-c_{12}c_{23}s_{13}e^{i\delta} &
-c_{12}s_{23}-s_{12}c_{23}s_{13}e^{i\delta} & c_{23}c_{13}} \nonumber
\label{eqone}
\]

\[
c_{ij} = \cos\theta_{ij} \quad , \quad s_{ij}=\sin\theta_{ij}
\]}

\noi (Our phase convention differs in sign from the PDG, but is more
 consistent with $V_{CKM}$).

Studies of atmospheric, $K2K$ and recent MINOS $\nu_\mu\to\nu_\mu$ 
disappearance indicate\cite{cipan} 

\addtocounter{equation}{1}
$$ 
 \Delta m^2_{32}=m^2_3-m^2_2 = \pm~2.6(3) \times 10^{-3} {\rm
eV}^2 \eqno{(2a)}
$$
$$
\sin^22\theta_{23}\simeq 1.0  \qquad\theta_{23}\simeq 45\pm{5^\circ} \eqno{(2b)}
$$ 

\noi The sign of $\Delta m^2_{32}$ is undetermined. For $m_3>m_2$,
normal ordering, neutrinoless double beta decay is highly suppressed,
while for $m_2>m_3$, inverted hierarchy, there is a chance that it could be 
 observable in the next generation of experiments. So,
determining the sign of $\Delta m^2_{32}$ is important. In the case
of $\theta_{23}$, maximal mixing, $\theta_{23}\simeq45^\circ$ is
favored. How close that angle is to $45^\circ$ and whether it is less than
 or greater than $45^\circ$ (currently only $sin^22\theta_{23}$ is determined)
is a key issue for model building. A very precise measurement is 
strongly warranted.

Solar neutrino and the Kamland reactor oscillation experiments 
indicate\cite{cipan}

\addtocounter{equation}{1}
$$
\Delta m^2_{21}=m^2_2-m^2_1=8\pm1\times10^{-5}{\rm eV}^2 \eqno{(3a)}
$$

$$
\sin^22\theta_{12}\simeq 0.84\pm 0.10,\quad\theta_{12}\simeq33^\circ\pm4^\circ 
\eqno{(3b)}
$$

\noi The angle $\theta_{12}$ is large but not maximal.

Within the 3 generation formalism, what remains to be determined are
the value of $\theta_{13}$,\\
 which is currently bounded\cite{cipan}

\be 
0\le\sin^22\theta_{13} \lsim0.14, 
\ee

\noi by reactor experiments, along with the phase, $\delta$, about
which nothing is currently known

\be
-180^\circ\le \delta < 180^\circ
\ee

\noi After those parameters are determined, one will have an intrinsic
measure of leptonic \\
CP violation via the Jarlskog invariant\cite{jarl}

\be
J_{CP} \equiv \frac18 \sin2\theta_{12}\sin2\theta_{13} \sin2\theta_{23}
\cos\theta_{13} \sin\delta.
\ee

\noi From the known angles ($\sin^22\theta_{12}\approx 0.8,$
 $\sin^22\theta_{23}\simeq1$)

\be
J_{CP}\simeq 0.23\sin\theta_{13}\sin\delta,
\ee

\noi which suggests it is potentially enormous in comparison with the
quark CKM matrix value 

\be
J^{CKM}_{CP} \simeq 3\pm1\times10^{-5}
\ee

\noi Besides determining the $\Delta m^2_{ij}$, their signs,
$\theta_{ij}$ and $\delta$ as precisely as possible, one would also
like to have precision redundancy in those studies which probes
deviations due to ``new physics'' such as sterile neutrino mixing,
extra dimensions, exotic neutrino interactions, etc.

\section{CP Violation}

The flavor changing oscillations $\nu_\mu\to\nu_e$ and
$\bar\nu_\mu\to\bar\nu_e$ have a very rich structure which includes
 CP violation. The oscillation probability is given by 3 important 
contributions as well as matter effects and smaller terms 
(which we neglect)\cite{marc,marcNuc}

\begin{equation}
 P(\nu_\mu\to\nu_e)=  P_I(\nu_\mu\to\nu_e) + P_{II}(\nu_\mu\to\nu_e) + P_{III}(\nu_\mu\to\nu_e)+ {\rm ~matter~}+{\rm smaller~~ terms} 
\end{equation}

\newpage
\begin{eqnarray}
P_I (\nu_\mu\to\nu_e) & = & \sin^2\theta_{23} \sin^22\theta_{13}   
\sin^2 \left(\frac{\Delta m^2_{31}L}{4E_\nu}\right) \\
P_{II}(\nu_\mu\to\nu_e) & = & \frac12 \sin2\theta_{12}             
\sin2\theta_{13} \sin2\theta_{23} \cos\theta_{13} \nonumber\\
 & & \sin
\left(\frac{\Delta m^2_{21}L}{2E_\nu}\right) 
 \times \left[\sin\delta\sin^2 \left(\frac{\Delta
m^2_{31}L}{4E_\nu}\right) \right. \nonumber\\
& & \left.
+ \cos\delta\sin \left(\frac{\Delta
m^2_{31}L}{4E_\nu}\right)  \cos \left(\frac{\Delta
m^2_{31}L}{4E_\nu}\right) \right]  \\
P_{III}(\nu_\mu\to\nu_e) & = & \sin^22\theta_{12}\cos^2 \theta_{13} 
\cos^2\theta_{23} \sin^2\left(\frac{\Delta m^2_{21}L}{4E_\nu}\right)
\end{eqnarray}
\noi while for $\bar\nu_\mu$, $\delta\to-\delta$ and matter effects 
change sign.  

The rich structure of $\nu_{\mu}\to\nu_{e}$ oscillations is nicely illustrated
in Figs.~1-4 for BNL-Homestake and Fermilab-Homestake distances.
Matter modifies the oscillation amplitudes and peak positions
(the effect is opposite for an inverted hierarchy), making it straight forward
to determine the sign of $\Delta m^2_{31}$ with only a $\nu_\mu$ beam.  Also, 
the effect of $\delta$ is important even for $\delta = 0$, no CP violation.
By measuring the $\nu_\mu$ oscillation probability as  function of
 a $ \frac{L}{E_\nu}$ over a broad rage, one can in principle measure all 
the parameters of neutrino oscillations with no degeneracies 
in $\delta$, $\theta_{23}$ and the mass hierarchy by a fit to Eq(9).  
For that reason, we favor\cite{marc,marcNuc,zp} using an on axis broad band neutrino 
beam  for  $0.5~GeV\le~E_\nu \leq 5~GeV$.  

Do we need to know the value of $\theta_{13}$ before we embark on measuring 
$\delta$?  Not really, since the degree of difficulty for measuring $\delta$ 
is to a large extent independent of   $\theta_{13}$ (unless it is very small)
 and the baseline distance (for $1200~km\lsim L\lsim 4000~km$ ) if we use 
the wide band beam. To see that feature, consider the CP violation asymmetry.

\be
A_{CP} \equiv
\frac{P(\nu_\mu\to\nu_e) -P(\bar\nu_\mu\to\bar\nu_e)}{P(\nu_\mu\to\nu_e)
+P(\bar\nu_\mu\to\bar\nu_e)} 
\ee

\noi It is given to leading order in $\Delta m^2_{21}$ (assuming
$\sin^22\theta_{13}$ is not too small) by 
 
\bea
A_{CP} & \simeq & \frac{\cos\theta_{23} \sin2\theta_{12}
\sin\delta}{\sin\theta_{23} \sin\theta_{13}} \left(\frac{\Delta
m^2_{21}L}{4E_\nu} \right) \nonumber \\
& & + {\rm matter~effects}
\eea

\noi For fixed $E_\nu$,
the asymmetry grows linearly with distance and increases as $\theta_{13}$ 
gets smaller. Of course $|A_{CP}|$ is bounded by 1; so, if it exceeds that
value, e.g.~if $\sin^22\theta_{13} \lsim0.003$, a breakdown in our 
assumption about the dominance of $P_I$ in the denominator of eq.(13) is
occurring.

The statistical figure of merit\cite{marc} is given by 

\be
F.O.M. = \left(\frac{\delta A_{CP}}{A_{CP}}\right)^{-2} =
\frac{A^2_{CP}N}{1-A^2_{CP}} 
\ee

\noi where $N$ is the total number of $\nu_\mu\to\nu_e +
\bar\nu_\mu\to\bar\nu_e$ events (properly normalized). Since $N$ falls
(roughly) as $\sin^2\theta_{13}$ and $A^2_{CP}\sim
1/\sin^2\theta_{13}$, we see that to a first approximation the F.O.M.
is independent of $\sin\theta_{13}$. Similarly, for a given $E_\nu$ the
neutrino flux and consequently $N$ falls as $1/L^2$ but that is
canceled by $L^2$ in $A^2_{CP}$. So, to a good approximation, our
ability to measure CP violation is insensitive to $L$(at oscillation max.)
 and the value of $\theta_{13}$ (if it is not too small).

\begin{figure}[h]
\begin{minipage}{18pc}
\includegraphics[width=18pc]{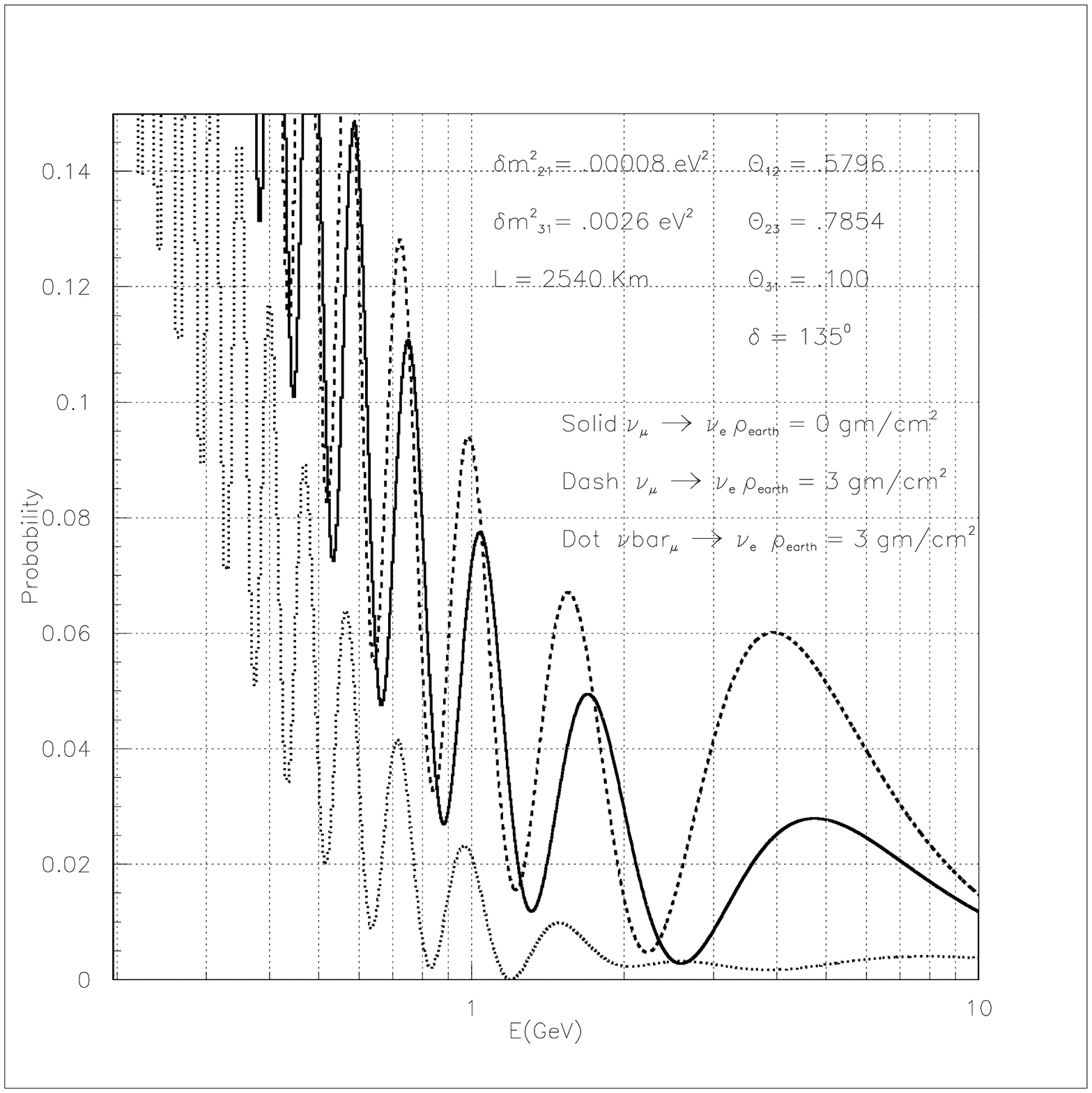}
\caption{\label{}}
\end{minipage}\hspace{2pc}
\begin{minipage}{18pc}
\includegraphics[width=18pc]{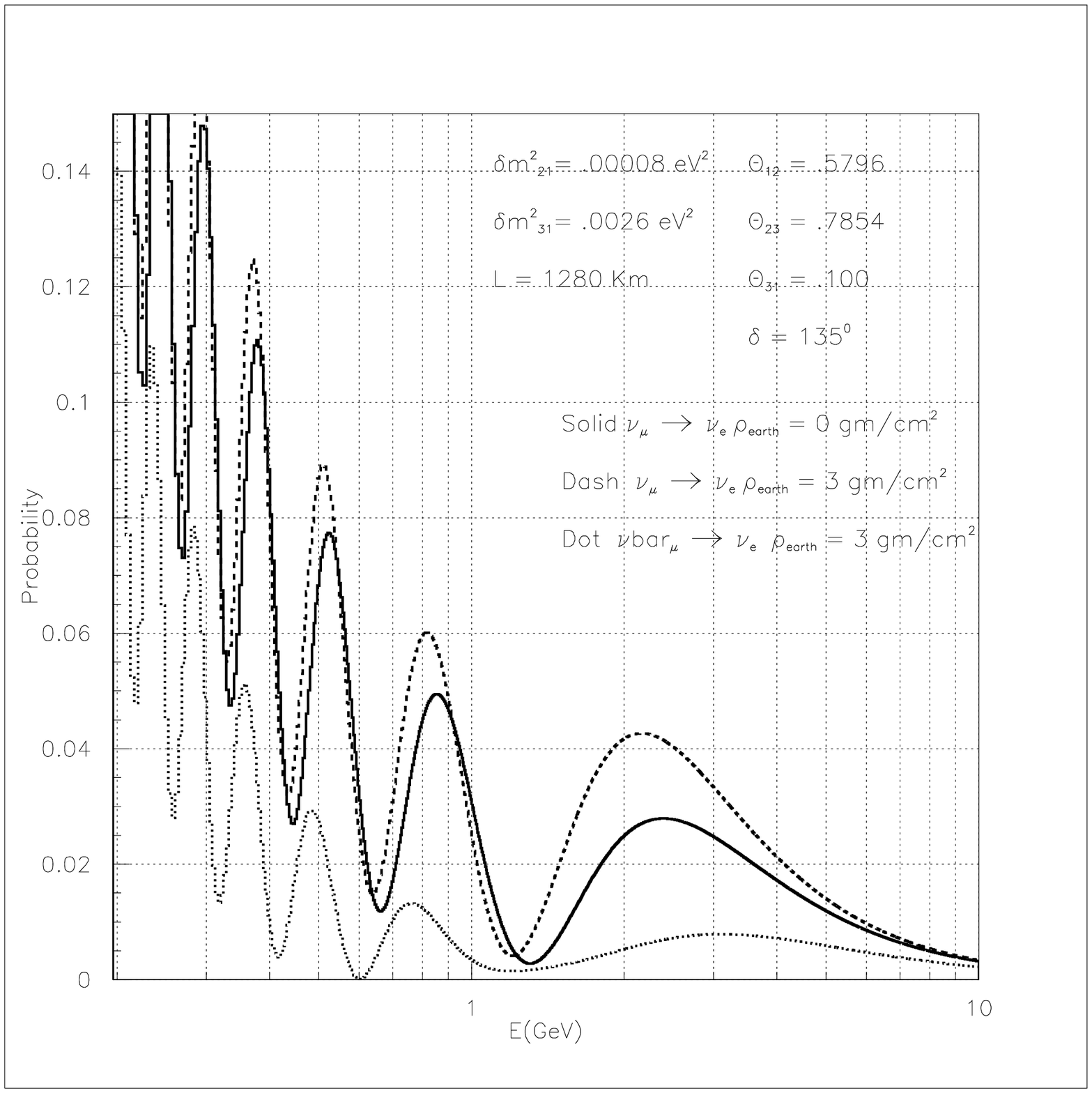}
\caption{\label{}  }
\end{minipage} 
\end{figure}

\begin{figure}[h]
\begin{minipage}{18pc}
\includegraphics[width=18pc]{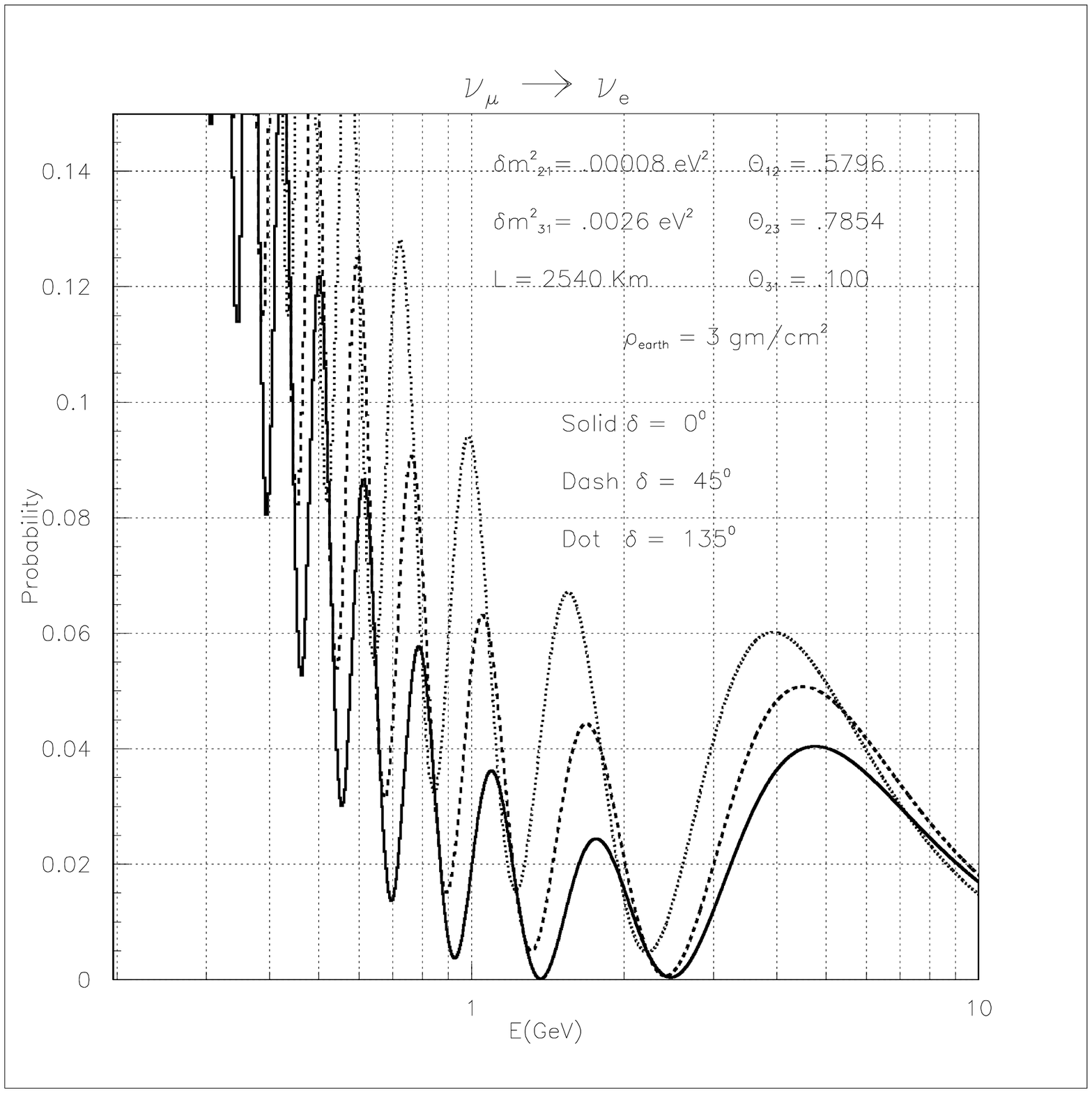}
\caption{\label{} }
\end{minipage}\hspace{2pc}%
\begin{minipage}{18pc}
\includegraphics[width=18pc]{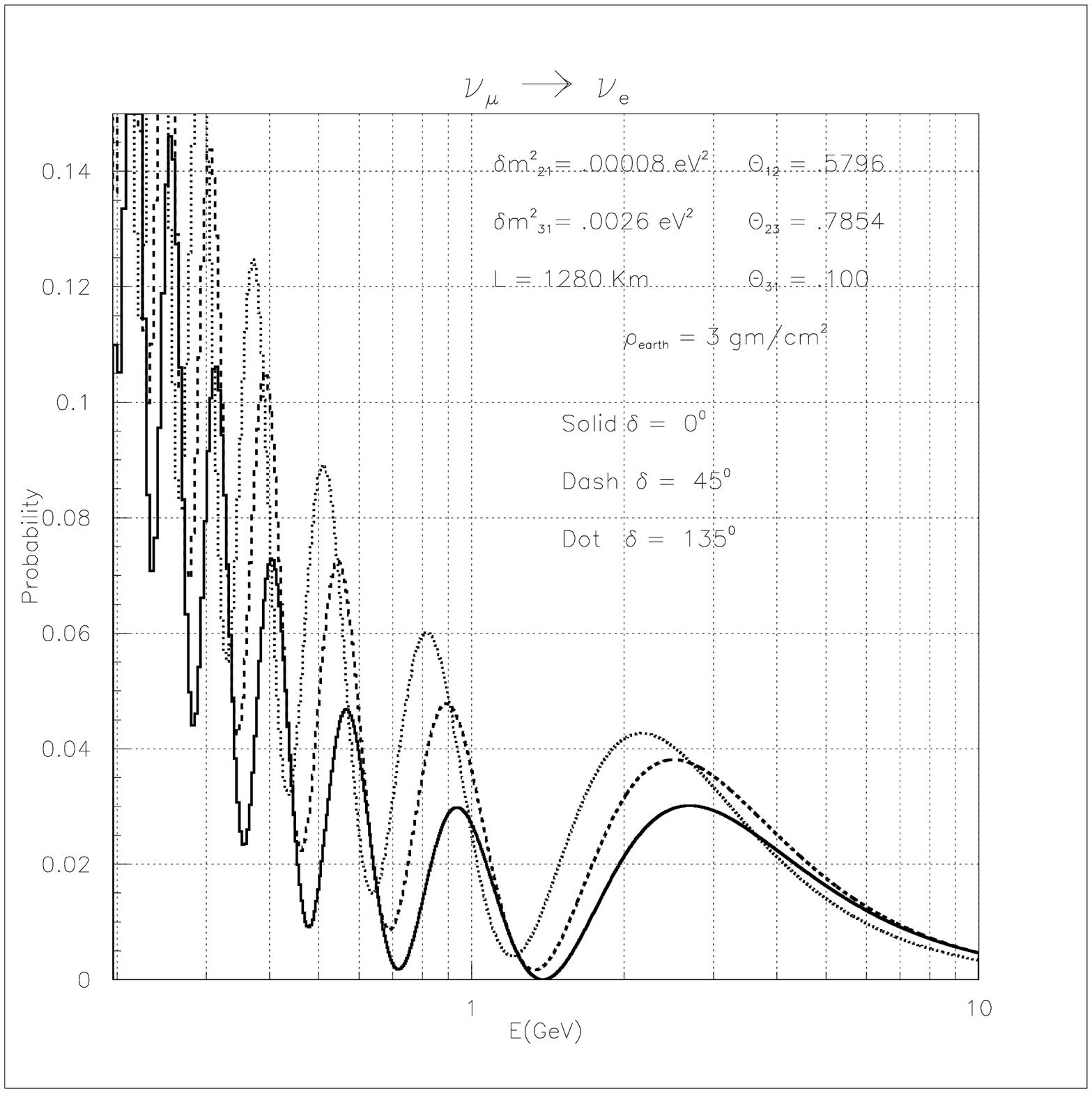}
\caption{\label{} }
\end{minipage} 
\smallskip
\begin{minipage}{37pc}

Fig 1-4. Neutrino oscillations, $\nu_\mu~ to~  \nu_e$, as a function of energy
for (Figs.~1~$\&$~3) BNL-Homestake (2540~km) and (Figs.~2~$\&$~4) 
FNAL - Homestake (1280~km). Effects of matter for neutrinos $\&$ 
antineutrinos relative to neutrinos with no matter are illustrated in
Figs.~1~$\&$~2, for $\delta=135^\circ$. A comparison of different 
phases $\delta=0, 45^\circ, 135^\circ$ is given in Figs.~3~$\&$~4. In all 
cases, we assume a normal mass
 hierarchy,~$\theta_{12}=0.5796$,
~$\theta_{23}=0.7854$~$\&$~$\theta_{13}=0.1~radians.$ 
\end{minipage} 
\end{figure}

Another way of seeing the insensitivity to $L$ in determining $\delta$
is to consider the 3 terms in eqs.~(10-12) separately. Each contributes to
$\nu_\mu\to\nu_e$ oscillations. The number of events from $P_I$ falls
as $1/L^2$ due to flux reduction while those from $P_{II}$ fall as
$1/L$ and from $P_{III}$ they are approximately constant (assuming $sin
\frac{\Delta m^2_{21}L}{4E_\nu} \sim \frac{\Delta m^2_{21}L}{4E_\nu})$.
Viewing $P_I$ and beam induced backgrounds (which also fall as $1/L^2$)
together as a total background for measuring $P_{II}$ and $P_{III}$, we
see that the determination of $P_{II}$ and therefore $\delta$ relative
to those backgrounds is independent of $L$ for fixed $E_\nu$ while the
$P_{III}$ signal to background increases linearly with $L$. So, longer
distances have some advantages for $P_{III}$. In addition, we see 
from eq. (11) that we can measure both $\sin \delta$ and $\cos \delta$ 
just by mapping out $\nu_\mu \to\nu_e$  oscillations 
(without antineutrinos) over a broad energy region. For those reasons, 
along with matter enhancement effects, larger $E_\nu$ high energy 
cross-sections, larger total neutrino flux etc.\ we advocate a wide band 
neutrino beam (on axis) $0.5\lsim E_\nu\lsim 5$ GeV and a large detector at
$1200-4000 km$ for the measurement of $\delta$. Our study of that idea
has shown many added benefits from the very long distance and broad band
beam. Indeed, in principle it allows measurement of $\Delta m^2_{31}$,
$\Delta m^2_{21}$, $sign~\Delta m^2_{31}$ $\sin^22\theta_{12}$,
$\sin^22\theta_{13}$, $\sin^22\theta_{23}$ and $\delta$ with
outstanding to good precision in one experiment, possibly with only
$\nu_\mu$ running (i.e.\ no $\bar\nu_\mu$). The basic features of that
proposal\cite{bnl} and some of its advantages are outlined below using 
a BNL-Homestake baseline, but first we explain why a conventional horn 
focused neutrino beam is currently the only viable way to explore 
leptonic CP violation.

\section{Other Intense Neutrino Beams\cite{zp}}

\subsection{Neutrino Superbeams:}

By definition, a neutrino superbeam would require a 4MW or 
more proton driver. Such a facility would deliver 4 times as much 
neutrino flux as a more conventional 1MW source. However, because of 
heat and increased radiation loads it would require liquid targets, 
robotic handling and special focusing horns or solenoids. The engineering 
requirements for 4MW are much more demanding, requiring significant R\&D to 
be realized\cite{zp}.  The cost for such a facility would be much higher 
than the more conventional 1MW proton driver and horn envisioned above. 
Preliminary discussions of 4MW sources for neutrino superbeams and their
anticipated oscillation studies are\cite{itow,autin}
${\rm JPARC~(Phase~II)}\to {\rm Hyper~K~}(1000{\rm~kton~}H_20), L=295km$,
${\rm CERN~(Super~linac)}\to {\rm Frejus~}(1000{\rm kton~}H_20), L=130km$.
Because of the relatively short baseline distances, those proposals would
employ only low energy neutrino flux $E_\nu<1$ GeV for their
 oscillation studies. That corresponds to only a fraction of the potentially 
available neutrino flux and the cross-section is lower.  To compensate, 
they must  employ enormous detectors (1000 Kton), a more powerful source, 
and long running time.  We have argued that it is much more cost
effective and richer in physics to use a wide band beam of higher
energy neutrinos and a much longer detector baseline 
distance\cite{marc,marcNuc,zp,bnl}.

\subsection{Neutrino Factory\cite{eleven}}

 Starting with an intense proton beam on target, the neutrino factory
concept envisions capturing the $\mu^{\pm}$ from $\pi^\pm\to\mu^\pm\nu$
decays, cooling them and then accelerating them to 20--50 GeV\null. At
that point they are placed in a storage ring with long straight
sections where the decays $\mu^+\to e^+\nu_e\bar\nu_\mu$ or $\mu^-\to
e^-\bar\nu_e\nu_\mu$ produce clean fluxes of high energy neutrinos
with $<E_\nu>\simeq 0.7$--$0.8E_\mu$. Neutrino factories are expected
to yield about 0.03$\nu_\mu$/proton; i.e.\ about 1/5  the flux of a
conventional horn focused neutrino beam. The neutrino factories
advantage (if it can be utilized) is the higher energy\cite{eleven}. 
The beam solid angle will scale as~$\sim 1/E^2_\mu$ and deep-inelastic
 cross-sections grow as $E_\nu$. Hence, at fixed distance one can 
gain $\sim E^3_\mu$ in event rate. However, in the case 
of oscillation studies, higher energies demand longer distance 
requirements for a fixed $L/E$ and a flux fall-off by $1/L^2$. 
That means, for $E_\nu\simeq20$ GeV to sit at the first oscillation peak
requires a detector at 12,000 km which is not possible. Hence, neutrino
factories must do their studies primarily at shorter distances
($\sim3000$km) where the first oscillation is only fractional. For
measuring $\theta_{13}$, the relative nearness is actually an advantage, 
but it is a drawback for CP violation studies which are optimized at
oscillation peaks. If $\theta_{13}$ is extremely small,
$\sin^22\theta_{13}\lsim0.003$, Neutrino Factories may be our best  hope
for  measuring it. However, in that case, CP violation and the phase
$\delta$ will be difficult to determine with such a facility.

\subsection{Beta Beam}

The interesting possibility of producing intense $\nu_e$ or $\bar\nu_e$ beams
 from nuclear beta decays was originally suggested by P.~Zucchelli\cite{zuc}.
It is particularly well matched to CERN's radioactive beams capabilities and 
accelerator complex. To be competitive with other intense neutrino facilities,
the radioactive nuclei must be copiously produced $\gsim$ $10^{13}/sec$, 
cooled, accelerated to $\gamma\simeq 100$ and kept in a large storage ring 
(with a long straight section) where a highly collimated  $\nu_e$ or
 $\bar\nu_e$ beams is produced by the decay
 $N\rightarrow N' e\bar\nu_e$.  Such a feat is extremely 
challenging, but the resulting beam has some very attractive features.  
It is absolutely clean, containing pure  $\nu_e$ or $\bar\nu_e$ with a 
precisely calculable energy spectrum.  Unlike the neutrino factory, it 
does not require a magnetized detector; so, a very large $H_2O$ Detector 
can be used. The neutrino energy spectrum is relatively low but broad, 
which are favorable characteristics for studying CP violation and 
measuring $\delta$. On the negative side, the flux is limited 
to $O(10^{18}\nu/yr)$ and the $\nu_{\mu}$ appearance cross-section is 
small.  CP violation studies lack statistics but may be marginally 
viable because of the potentially tiny backgrounds.

\section{BNL-HOMESTAKE NEUTRINO OSCILLATION EXPERIMENT}
 
We have written a white\cite{bnl} paper and had several follow-up 
studies extolling the virtues of a very long baseline
 BNL-Homestake (2540~km) neutrino oscillation experiment. (Actually,
 any distance\cite{marc} from about 1200--4000~km will do.) Its basic 
requirements are: 1) A conventional horn focused intense $\nu_\mu$ beam 
using an upgraded 1-2~MW AGS proton beam on a standard target. The cost and  
technical requirements \cite{bnl} needed for the upgrade are modest in 
comparison with ideas for 4MW superbeam or neutrino-factory sources 
described above. The resulting neutrino beam (on axis at $0^\circ$) would 
be broad band, 0.5 GeV${}\lsim E_\nu\lsim5$ GeV, peaking near 1.5 GeV\null.  
2) The detector\cite{detec} would be about a 250-- 500 kton
water cherenkov detector and would likely be somewhat modular in design.
This is again modest (about half the cost) in comparison with the 1000 kton 
behemoth detectors being considered by others. To reconstruct the neutrino
energy on an event by event basis and reject $\pi^0$ background, we would 
primarily  use quasi-elastic events 
$\nu_e n\to e^-p$ in the analysis. They
represent less than 1/4 of all neutrino events; therefore, a detector
with better resolution and acceptance such as liquid Argon or
Scintilator could be smaller, in principle, of order 100--200 kton
by using a larger fraction of events to do the job. 3) The run time
would be about $5\times10^7$ sec with a $\nu_\mu$ beam. Two types of
oscillation measurements would be made $\nu_\mu\to\nu_\mu$
disappearance and $\nu_\mu\to\nu_e$ appearance. At a later time
$\bar\nu_\mu$ studies might be carried out; however, they may not be
necessary because the wide band beam allows sensitivity to all neutrino
oscillation parameters, even $\delta$, without actually measuring a CP
violating effect such as $A_{CP}$ directly. Instead a fit is done to the
data assuming 3 generation mixing.

Because of the long distance and broad band beam, many physics studies
are possible. The measurement of $\nu_\mu\to\nu_\mu$ disappearance

\begin{equation}
P(\nu_\mu\to\nu_\mu) = 1-\sin^22\theta_{23}\sin^2 \left(\frac{\Delta
m^2_{31}L}{4E_\nu}\right)+{\rm~smaller~terms} 
\end{equation}
\noi over the range $0.5\le E_\nu \le 5$ GeV would be sensitive to 3 or 4
oscillation cycles \cite{bnl}. 
Such measurements would determine $\sin^22\theta_{23}$ and $\Delta
m^2_{31}$ to better than $\pm1\%$ statistically. Such a study will tell us if
$\theta_{23}\simeq 45^\circ$ to within about $\pm2^\circ$. Also, by
comparing values of $\Delta m^2_{31}$ obtained at different $E_\nu$,
one can search for indications of ``new physics''.

The study of $\nu_\mu\to\nu_e$ oscillations can be divided into three
domains: 1) High Energy, 3 GeV${}\le E_\nu\lsim5$ GeV, 2) Intermediate
Energy, 1 GeV${}\le E_\nu\le 3$ GeV and 3) Low Energy, $E_\nu\lsim1$
GeV\null. Roughly speaking, the high energy $\nu_e$ events will be
matter enhanced (suppressed) for the normal (inverted) mass hierarchy.
The effect is very pronounced (see Figs.~1~$\&$~2), making a determination of
the sign of $\Delta m^2_{31}$ relatively easy (for
$\sin^22\theta_{13}\ge0.01$) and allowing for a good measurement or
bound on $\theta_{13}$ (via $P_I$) which is better than any other
proposed experiment \cite{bnl}. Intermediate energy events will
measure both $\sin\delta$ and $\cos \delta$ via $P_{II}$. In that way
we expect $\delta$ to be determined to within $\pm15^\circ$ independent of its
value with no ambiguity \cite{bnl} (again assuming
$\sin^22\theta_{13}\gsim0.01$). That type of $\delta$ determination is
more robust and statistically more powerful than $A_{CP}$.  Note, that
the energy peaks are also displaced by matter effects. Their positions
can in principle be used to determine the sign of $\Delta m^2_{31}$. 
(see fig.~1.) Finally, the low energy $\nu_e$ events will determine the 
combination 
$\Delta m^2_{21}\sin2\theta_{12}$ to about $\pm5\%$ via $P_{III}$. Altogether,
this single experiment will measure or constrain all parameters of 3
generation leptonic mixing with unprecedented sensitivity and without 
parameter degeneracies. It would put
leptonic mixing on about the same level of precision as quark mixing.
Specific detains of detector optimization and running strategy still
need to be ironed out, but the basic idea of determining all
oscillation parameters via one experiment is very compelling.
We also note, a Fermilab-Homestake (1280~km) and wideband beam experiment 
would exhibit less dramatic effects (see Fig.~2),but would have about 4 
times the statistics because of the shorter distance.  Overall, it would 
have similar discovery potential. Figs.~3~$\&$~4 illustrate the 
dependence on the phase $\delta$ for BNL and Fermilab distances.

\section{OUTLOOK}
It appears that the combination of intense conventional wide band
$\nu_\mu$ beam, powered by a 1-2 MW proton accelerator, large detector 
and very long baseline provides an opportunity to measure 
$\Delta m^2_{31}$, sign $\Delta m^2_{31}$,
$\Delta m^2_{21}$, all $\theta_{ij}$ and $\delta$ with good to high
precision. The intense proton source required for this effort is a
straightforward upgrade of the AGS or Fermilab Main Injector\null. 
The large detector (= 500 kton $H_2O$ or its equivalent) could be 
sited  at either of the national underground lab sites being 
considered (Homestake or Henderson). It would
also search for proton decay, supernova, atmospheric neutrinos etc.
to unprecedented levels. The facility would probably be at the
forefront of particle physics research for 50 years or more. Of course, 
proton souces at JPARC or CERN are also options for such a long baseline 
effort.

What remains to be done? Detector R\&D to reject backgrounds such as
$\pi^0$ and reduce the cost are needed. An underground lab
site needs to be developed and the horn generated wide band beam flux 
should be optimized. After the first phase of $\nu_\mu$ is completed, 
one might run $\bar\nu_\mu$ for a few years if one wants to actually 
observe CP violation (rather than just a determination of $\delta$) or 
if an inverted mass hierarchy turns out to be correct. During that 
time further upgrades of the AGS or Main Injector to 2MW or more might 
be appropriate.

The strategy for long baseline neutrino oscillations outlined here is
based on novel concepts: broad band beam, very long distance and large
detector. It is bold, ambitious and doable. The opportunity is within
our community's grasp and should be seized.

\end{document}